\documentclass[conference]{IEEEtran}
\ifCLASSINFOpdf
\else
\fi

\usepackage{multirow}
\usepackage{graphicx}
\usepackage{epstopdf}
\usepackage{cite}
\usepackage{subfigure}
\usepackage{verbatim}

\usepackage[noend]{algpseudocode}

\usepackage{algorithmicx,algorithm}

\usepackage{CJK}

\renewcommand{\thetable}{\alph{table}}

\begin{document}
\graphicspath{{fig/}}

%

\title{A Coordinated View of Cyberspace}

\author{
\IEEEauthorblockN{
Congcong Miao\IEEEauthorrefmark{1}\IEEEauthorrefmark{2}\IEEEauthorrefmark{3},
Jilong Wang\IEEEauthorrefmark{2}\IEEEauthorrefmark{3},
Shuying Zhuang\IEEEauthorrefmark{2}\IEEEauthorrefmark{3},
Changqing An\IEEEauthorrefmark{2}\IEEEauthorrefmark{3}
}

\IEEEauthorblockA{\IEEEauthorrefmark{1}Department of Computer Science and Technology, Tsinghua University}
\IEEEauthorblockA{\IEEEauthorrefmark{2} Institute for Network Sciences and Cyberspace, Tsinghua University}
\IEEEauthorblockA{\IEEEauthorrefmark{3}Beijing National Research Center for Information Science and Technology}
\IEEEauthorblockA{Email: mcc15@mails.tsinghua.edu.cn, wjl@cernet.edu.cn}
}


%


\maketitle

\begin{abstract}

Cyberspace, created by communication technologies especially the Internet, is a virtual space where people can easily communicate with others regardless of geographic distance. Due to its convenience, cyberspace has been widely applied in people's daily life and it is regarded as a new space, paralleled to the geographic space. The development of cyberspace has gradually attract the attention in academic researches. Unfortunately, the complexity and multi-dimension of cyberspace pose great challenges and researchers even have not defined its basic coordinate system. Therefore, many prior efforts have tried to demonstrate the cyberspace using geographic coordinate system (GCS) and geographic map (GeoMap). It is useful for sometimes to demonstrate the virtual space with aid of the real world maps. However, it is not suitable for the study of cyberspace in many scenarios, such as network management and network security. To best of our knowledge, this paper is the first work to study the basic coordinate system of cyberspace. Firstly, we search for the space coordinate as the basic space dimension based on a number of alternative choices (such as IP address, AS numbers, etc.), and then construct \emph{Cyberspace Coordination System} (CyberCS) based on the space filling curves or combination of space coordinates. Finally, we validate the effectiveness of our CyberCS on the cyberspace maps (CyberMap) with various real world network traces. In addition, a combination of different maps provides multiple views of cyberspace and enables visual correlation from the macroscopic perspective.

\end{abstract}

%
\IEEEpeerreviewmaketitle

\section{Introduction}
\label{sec:introduction}
Cyberspace, created by communication technologies especially the Internet, is a virtual space where people can access, interact and communicate through physical devices independent of time and space. It is a digital space and has no actual physical location. Therefore, people can easily communicate in cyberspace regardless of the limitation of geographical distance. Due to its convenience, cyberspace has been widely applied in people's daily life and it gradually becomes an indispensable space, paralleled to geographic space. However, the virtuality, complexity and multi-dimension \cite{jinhua2013research} of cyberspace pose great challenge and it is hard to define its basic coordinate system to now.

The development of cyberspace has gradually drawn more attention in academic researches and many prior efforts have tried to demonstrate cyberspace. Jiang et al. \cite{jiang1997cybermap} take the first step into mapping cyberspace and divide the cyberspace into physical dimension and information dimension. Later on, researchers are motivated by the geographic coordinate system (GCS) and geographic map (GeoMap), so that they try to demonstrate cyberspace in geographic space, such as traffic visualization \cite{caida}, anycast geolocation \cite{cicalese2015characterizing}, worm propagation \cite{trullols2015worm}, etc. There is no doubt that both GCS and GeoMap are very useful for sometimes to demonstrate cyberspace with the aid of real world map. However, it is limited since it only demonstrate the geographic characteristic of cyberspace while it can not directly demonstrate cyberspace. In other words, it present the geographical projection of cyberspace. Taking a general cyberspace case as an example, visualizing network attack in geographic space is easy to get the location information, however, it is hard to get the deeper insights of network information which the viewer are more concerned, such as IP information.

Other works try to visualize cyberspace with topology coordinate system (TopoCS) and topology map (TopoMap) which seems an intuitive way to demonstrate cyberspace. The key idea of TopoCS is to model the node on the network while referring to geometric coordinate system to construct TopoCS, and then characterize the position of each node. By this way, the topological relation and distance of node in cyberspace is calculated and depicted on TopoMap. Note, cyberspace is created by communication technologies that the node and link are the main components of cyberspace. There are a lot of works focusing on the TopoMap such as WSP \cite{zhu2012wsp}, DMFSGD \cite{liao2013dmfsgd} and DISCS \cite{cheng2017discs}. These works present the real situation of cyberspace that the length of link represents the distance and the node represents the location. However, the topology coordinate system changes frequently due to the network distance changes and topology changes. We can not use a self-unstable system as a cyberspace coordinate system.

The traditional well known coordinate systems and maps show their limitation to demonstrate cyberspace and the complex and multi-dimensional cyberspace poses great challenges. However, cyberspace is in great need of its own coordinate system to be a guide to other studies of cyberspace. We take this opportunity and take first step into demonstrating cyberspace with cyberspace coordinate system (CybderCS).  For simplicity, we only consider the special part of cyberspace which is combined of the Internet. The main contributions of our work are as follows. We firstly search for the space coordinate as the basic space dimension based on a number of alternative choices (such as IP address, AS number, Port number, MAC address, etc), and then construct CyberCS based on space filling curves or combination of space coordinates. We validate the power of our CyberCS on the CyberMap with various real world network traces. The results show that the CyberCS and CyberMap present deeper insights into the cyberspace that different parts and degrees of cyberspace are efficiently visualized. In addition, similar to the Fourier Transform, a combination of different maps provides multiple views of cyberspace and enables visual correlation from the macroscopic perspective.

The rest of this paper is organized as follows. Section \uppercase\expandafter{\romannumeral2} reviews in detail the related works in traditional well known coordinate systems. In section \uppercase\expandafter{\romannumeral3}, we deeply analyze and discuss the reason to choose the coordinates. Section \uppercase\expandafter{\romannumeral4} provides the detailed construction of CyberCS based on the space filling curves or combination of space coordinates and section \uppercase\expandafter{\romannumeral5} presents the application of proposed CyberCS on CyberMap with real world netwrok traces. We present a comparison with different maps in section \uppercase\expandafter{\romannumeral6} and finally section \uppercase\expandafter{\romannumeral7} outlines our conclusions by reminding its main contributions.

\section{Related work}
The development of cyberspace has gradually drawn more attention in academic research, resulting in a number of works focusing on representing and visualizing cyberspace. The main concept of these work on mapping cyberspace is to project the cyberspace on the current coordinate system or establish a new coordinate system with aid of the geometric coordinate system. The related works are presented from two aspects, the former is based on GeoMap with GCS, while the latter is based on TopoMap with TopoCS.

\textbf{GeoMap:} Demonstrating cyberspace in GeoMap is the most common and understandable way with latitude and longitude as the basic coordinates. Since the viewer is used to the GeoMap, the researchers try to present the detailed geographic location of cyberspace which is easy for the viewer to understand. There are many works on visualizing various parts of cyberspace. Castro et al. \cite{castro2010understanding} focus on the DNS service in cyberspace. They measures the geographic distribution of clients querying the root DNS server and then visualize the geographic workload distribution of DNS clients for each anycast instance. Another work by CAIDA develops a geographic visualization tool with GeoMap called Cuttlefish \cite{caida2} which produces animated GIFs that reveal the interplay between the diurnal and geographical patterns of cyberspace events. It uses two cases, one is Japanese ISP traffic to show the correlation between the residential broadband customer traffic through the Japanese ISP and the time of day in Japan, the other is to show the propagation rate at which the Witty Internet Worm infected hosts in geographic world. King et al. \cite{king2014a} also present two cyberspace events in GeoMap to monitor the evolution of events over time, one is the botnet-orchestrated stealth scan, the other is the government mandated Internet blackout in Egypt. Demonstrating cyberspace in GeoMap presents an geographic projection of cyberspace. However, GeoMap for cyberspace falls short of presenting the deeper insights of network information which the viewer are more concerned, such as IP information.

\textbf{TopoMap:} TopoMap is widely used to demonstrating the topology connection of cyberspace. The key idea of TopoMap is to model the cyberspace in geometric space by establishing TopoCS. Specifically, we calculate the network distance of node in cyberspace to determine suitable TopoCS and then map the position of each node based on the distance. There are a lot of works focusing on how to establish the TopoCS, including landmarks-based TopoCS and decentralized TopoCS. Landmarks-based TopoCS selects a set of fixed hosts as references to compute its own coordinate system while other hosts performs latency measurement with fixed hosts to find a position in TopoCS. WSP \cite{zhu2012wsp} is a Web service positioning framework to make response time prediction for Web services. There are numbers of fixed points deployed in the network to construct TopoCS. Contrast to landmarks-based approach, decentralized TopoCS neither relies on explicitly well-known hosts nor requires any host in the system to act as a reference node. DMFSGD \cite{liao2013dmfsgd} only requires each node to collect local measurements, with neither explicit matrix constructions nor fixed nodes as landmarks. However, it provides scalability and high accuracy. DISCS \cite{cheng2017discs} establishes robust nonnegative matrix completion method to eliminate measurement error to improve prediction accuracy. TopoMap present the real situation of cyberspace that the length of link represents the distance and the node represents the location. However, the network topology changes frequently due to the network distance changes. We can not use a self-unstable system as a cyberspace coordinate system.

\section{Cyberspace Coordinate}
The GeoMap with geographic coordinate system (latitude and longitude) and TopoMap with network coordinate system are not effective to visualize cyberspace. However, it is in a great need of its own coordinate system for further demonstrating cyberspace. The virtual and complex cyberspace presents several big challenges: (1) How many space dimensions will cyberspace has? (2) What is the basic 2-d or 3-d dimension definition of cyberspace like the physical world? (3) Especially, are there any stable numbering systems of cyberspace can be considering as a basic space dimension like latitude and longitude?  Motivated by the geographic coordinate system (latitude and longitude) which is stable numbering system, we also look for the stable numbering system (coordination) in cyberspace as the basic space dimension to construct the cyberspace coordinate system. With deep understanding of cyberspace, we observes a number of alternative choices such as IP address space, Autonomous System (AS) number space, MAC address space, domain name space and port number space. These coordinations are stable and widely adopted that almost all objects in cyberspace possess them as identifiers so that they are able to project the cyberspace in its own space. We are discussing each coordination in the following.

The GeoMap with geographic coordinate system and TopoMap with topology coordinate system show their limitation to demonstrate cyberspace. However, cyberspace is in great need of its own coordinate system to be a guide to other studies of cyberspace. The virtual, complex and multi-dimensional cyberspace presents several big challenges: (1) How many space dimensions will cyberspace has? (2) What is the basic 2-d or 3-d dimension definition of cyberspace like the physical world? (3) Especially, are there any stable numbering systems of cyberspace can be considering as a basic space dimension like latitude and longitude in GCS?  

Motivated by GCS which is composed of stable numbering system (i.e., latitude from $-180^{o}$ to $180^{o}$ and longitude $-90^{o}$ to $90^{o}$), we also look for the stable numbering system as a coordinate in cyberspace to construct the CyberCS. The coordinate should also be able to represent the basic space dimension of cyberspace. With deep understanding of cyberspace, we observes a number of alternative choices such as IP address space, Autonomous System (AS) space, MAC address space, domain name space and port space, etc. Firstly, They are stable numbering systems. More importantly, these coordinations are widely used in cyberspace as identifiers so that they are able to project the cyberspace in its own space. We will discuss each coordinate detailedly in the following.

\textbf{IP address:}
The IP address is a stable numbering system which is composed of a fixed bit number. The total number of IPv4 is 2$^{32}$ while the total number of IPv6 is 2$^{128}$. The IP address is the unique fingerprint assigned to object which is regarded as very importance in cyberspace. It serves two primary functions. On the one hand, it is used as a network interface identification that allows host to send and receive information and communicate with others in cyberspace. On the other hand, it provides the location of an object, similar to a physical address (longitude and latitude) in geographic space. All behaviors and information interactions in network need to be implemented based on IP addresses. Since the number of IP address doesn't change with network status and the importance of IP address in cyberspace, projecting the cyberspace in IP address space is an effective way to demonstrate cyberspace which could provide the valuable IP information.

\textbf{Port :}
An port number is composed of a fix-length 16-bit binary number. The total number of port is stable, numbering form 0 to 65535. A network port serves as a logical communication endpoint to allow different services on the same carrier to share network resources simultaneously. The port number is used to identify the specific services running on that carrier. Generally, the port number is often come up with an IP address when establishing a network connection to represent a specific service on cyberspace. An IP address is the network address of a carrier which projects the cyberspace IP address space, while port number is the logic address of a specific service which further projects the cyberspace in port number space. Similar to the GCS which projects the geographic space into in latitude and longitude, the combination of IP address and port number is an effective way to help the viewer to take further step into cyberspace because it could provide the various application layer information.



\textbf{AS number:} The AS number is composed of a 16-bit binary number with the total number of 2$^{16}$ and the AS number is also a stable numbering system. AS is defined for routing policy on the Internet and combined of a collection of IP address under the control of the network operator. Each AS contains a set of IP addresses and the relationship between IP address and AS are operated by RIRs. Therefore, AS is also regarded as the location of aggregated objects in cyberspace. Projecting the cyberspace into AS space provide the aggregated characteristics of IP address space. It is also an effective way to demonstrate cyberspace if the viewer want to visualize the AS level information of cyberspace such as the AS topology.

\textbf{MAC Address:} MAC address, defined as Media Access Control Address, is a unique identifier of network interfaces through a physical network segment. In other words, it's an identifier of hardware that uses Ethernet, which can also be referred as physical address or hardware address. Since the MAC address is the stable numbering system that is composed of 12 characters, so it could be used for the coordination of cyberspace. Furthermore, the cyberspace is created by the physical network resource with MAC address, so that we can project the cyberspace into MAC address space which is traced into each physical host.

\textbf{Domain Name:} Domain name is alphabetic which is easier to remember. For example, the domain name has a formed name e.g. www.apple.com, which is the identification of Apple company. The last part of the address, called top-level domain, shows the type of organizations(e.g. .com, .gov) or locations(e.g. .cn, .uk). Domain Name System(DNS) is a distributed database with IP address and maintains the domain name, which provides translation between domain name and IP address, e.g. 17.16.0.1 = www.apple.com. Domain name is a stable numbering system which is not change with network status, however, it is impossible to enumerate because the length of domain name can be variable. Projecting the cyberspace into domain name space only provide the detailed web information of cyberspace.

We have discussed a number of alternative choices which could be used as the coordinates of cyberspace. Each coordinate is a candidate to construct the CyberCS. However, projecting the cyberspace into MAC address space and domain name space is not very effective and may result in poor visualization of cyberspace. The former may lead to the sparse visualization because majority of MAC addresses are not connected to the Internet while the latter only provide the detailed web information which is regarded as a small part of cyberspace. As for IP address space, port space and AS space which can be regarded as the location of object in cyberspace, they seems more proper to demonstrate cyberspace. We choose these coordinates to design CyberCS in the next section.

\section{CyberCS}
We have discussed a number of stable numbering systems, i.e., IP address, port and AS, which could be used as the coordinate of cyberspace. In this section, we describe in detail the design of our cyberspace coordinate system (CyberCS) based on these coordinates. Similar to the GCS with 2-D horizontal coordinates or 3-D Cartesian coordinates to demonstrate the geographic space, we design 3 types of CyberCS to demonstrate cyberspace. Each coordinate system could be used to demonstrate cyberspace in different scenarios.


\subsection{CyberCS based on IP address}
The IP address is regarded as the location of an object in cyberspace which is a proper coordinate to demonstrate cyberspace. However, directly projecting the cyberspace into 1-D IP address space may lead to poor visualization of cyberspace. Specifically, the total unique IP address is 2$^{32}$ (IPv4) which means the length of the coordinates is more than 4 billions. It is not intuitive to represent cyberspace and hard for the viewer to understand. Therefore, we design 2-D IP coordinates with the space filling curves algorithm which could extrapolate the IP number from one dimension to two dimensions.

According to \cite{asano1997space}, there are several space filling curves, including z-order curve, the Gray code curve and Hilbert curve which could convert one-dimension sequential data into two dimensions. By analyzing the clustering properties of the above space filling curves, the Hilbert curve is proved achieving the best clustering properties \cite{moon2001analysis} - particularly the notion of ordering and closeness to sequential nodes. The Hilbert curve maintains locality of the data on the curve which means that the order of data in one dimension will still be ordered in the same way in two dimension. Another property is that it visits every lattice point in a square with side length a power of two which is especially suitable for the IP address since the length of IP address occurs in powers of two. 

\begin{algorithm} [t] \small
	\caption{Hilbert space filling algorithm}\label{alg1}
	\hspace*{0.02in} {\bf Input:} 
	$n$
	\begin{algorithmic}[1]
		\State {\textbf{Initialization:}} $x,y,x_i,y_j \gets 0$, $x_j,y_i \gets 1$
		\Function {$Hilbert$}{$x$, $y$, $x_i$, $x_j$, $y_i$, $y_j$, $n$}
		\If {$n == 0$}
		\State $Px \gets x+(x_i+y_i)/2$, $Py \gets y+(x_j+y_j)/2$
		\State \textbf{LineTo}($P_x$, $P_y$);
		\Else
		\State{\Call{$Hilbert$}{$x$, $y$, $y_i/2$, $y_j/2$, $x_i/2$, $x_j/2$, $n-1$}};
		\State{\Call{$Hilbert$}{$x+x_i/2$, $y+x_j/2$, $x_i/2$, $x_j/2$, $y_i/2$, $y_j/2$, $n-1$}};
		\State{\Call{$Hilbert$}{$x+(x_i+y_i)/2$, $y+(x_j+y_j)/2$, $x_i/2$, $x_j/2$, $y_i/2$, $y_j/2$, $n-1$}};
		\State{\Call{$Hilbert$}{$x+x_i/2+y_i$, $y+x_j/2+y_j$, $-y_i/2$, $-y_j/2$, $-x_i/2$, $-x_j/2$, $n-1$}};
		\EndIf
		\EndFunction		
	\end{algorithmic} 
\end{algorithm}

\begin{figure}[t]
	\includegraphics[width=0.45\textwidth]{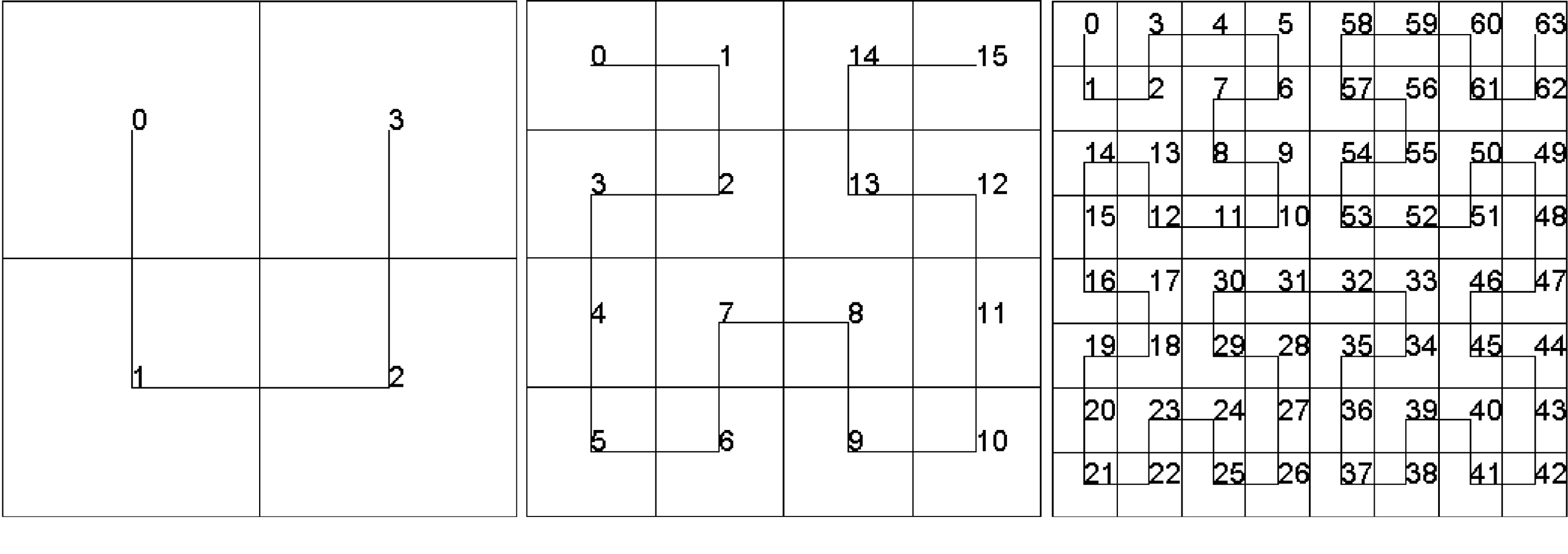}
	\centering
	\caption{The Hilbert curve with order 1, 2 and 3.}
	\label{fig_ipcensus} 
\end{figure}

Algorithm \ref{alg1} describes how to generate a Hilbert map. The input $n$ is the order of Hilbert curves, which is defined by the number of data in one dimension. For example, if the number of IPv4 is $2^{32}$, the order should be set 16 so that it could be mapped with $2^{16} * 2^{16}$ in two dimension. In each order $n$, there are six inputs that the first two define the coordinates of the current input point (x,y) and the next 4 values define two vectors to represent the direction of next point (line 2). If the order is 0 which means the maps in two dimensions becomes a point, so that we can generate the next point according to a certain directions and line to it (line 3-5). Other wise, we perform the arithmetic on the current point (x,y) for the current level of recursion with the fixed transformation converting each line to a smaller version of the original open square (line 7-10). For each recursion, we can get four smaller separate squares. Fig. \ref{fig_ipcensus} is the generated Hilbert curve with order 1,2 and 3. Based on the Hilbert curve, the IP address could be extrapolated from one dimension into two dimension to generate the 2-D IP coordinate system.

\begin{figure}[b] 
	\begin{minipage}[t]{0.22\textwidth} 
		\centering 
		\includegraphics[width=0.95\linewidth]{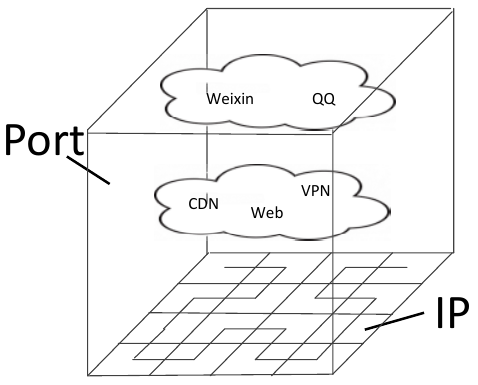} 
		\caption{CyberCS based on IP-Port} 
		\label{fig:side:a} 
	\end{minipage}%
	\begin{minipage}[t]{0.22\textwidth} 
		\centering 
		\includegraphics[width=0.95\linewidth]{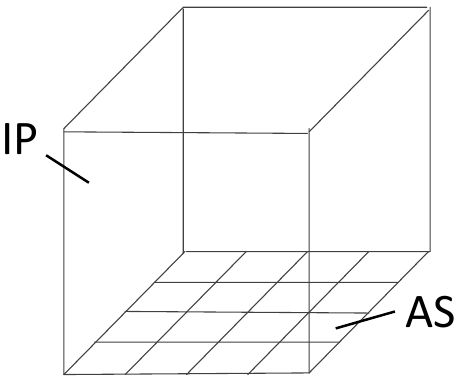} 
		\caption{CyberCS based on AS} 
		\label{fig:side:b} 
	\end{minipage} 
\end{figure}
\subsection{CyberCS based on IP-Port}
Although the basic 2-D IP coordinate system is able to project cyberspace in a unified way, it would be difficult to take further step into cyberspace and demonstrate more detailed view of cyberspace, which might be of great interest to people. To achieve this, we extend the basic two-dimensional IP drawn by the Hilbert curve mapping algorithm into a 3-D coordinates. In order to choose the suitable coordinate as z axis, we have throughly thought about the alternative choices discussed in Section \uppercase\expandafter{\romannumeral3}.

Based on our observation, we find that the port number is a very suitable choice to be the z axis because of three reasons. Firstly, it is a stable numbering system that the number of port does not change with the network status. Then, the coordination of the port number is orthogonal to the coordination of IP address which meet the basic requirement of the coordinate system. Most importantly, the port number ofter comes up with an IP address to describe the application layer of cyberspace, a more detailed perspective of cyberspace. Therefore, the viewer is able to get the application layer, e.g.. a process or a network service, of cyberspace. 

The 3-D IP-Port coordinates, as shown in Fig. \ref{fig:side:a}, could demonstrate a fine-grained activity of cyberspace. For example, we can get the detailed application running on each port, such as port 80 for web service and port 502 for modus application. Furthermore, the application level traffic running on each port can also be visualized on the IP-Port map, so that the network administrator could take further step into monitoring these applications for abnormal behavior detection.

\subsection{CyberCS based on AS}

The AS is combined of a collection of IP addresses so that the cyberspace could be also regarded as a combination of multiple autonomous systems (ASes). The 2-D IP coordinates and 3-D IP-Port coordinates could demonstrate the cyberspace in most scenarios. However, they failed in intuitively demonstrating the AS level activity of cyberspace since the assignment of IP address segments under an AS may be discontinuous, resulting in poor visualization of the IP address-based map. For example, the AS 4538 is combined of multiple IP blocks which are not aggregated in the IP coordinates, so it is not intuitive to represent the activity of AS4538 and hard for the viewer to understand in the IP coordinates. Therefore, we propose the AS coordinates to represent cyberspace.

Fig. \ref{fig:side:b} presents the detailed AS coordinate model. The AS is chosen as a basic coordinate and then extrapolated from one dimension into two dimension with spacing filling curves. The extrapolation curves could be z order curve, the Gray code curve and the Hilbert curve. In this paper, we choose the Hilbert curve as the spacing filling curves. Therefore, the 2-D plane is constructed to represent the AS information which is similar to the expression of national information by latitude and longitude in GCS with longitude and latitude. In order to enrich the AS map, we add the IP address as the third dimension, which is used to be as the attribute of AS. The detail is shown in \uppercase\expandafter{\romannumeral5}.

\section{CyberMap}
With the aid of CyberCS which could be widely used in cyberspace, we introduce the Cyberspace map (CyberMap) with Internet measurement data. For each type of CyberCS, we illustrate one application as an example to demonstrate the specific part of cyberspace and take cyberspace management and security as a background to show the power of cybermap.
\subsection{IP Map}
The Hilbert curve is used to extrapolate IP address from one dimension into two dimensions to get the 2-D IP coordinate system. Therefor, we are able to project the cyberspace into a 2-D plane to get the IP map. The IP map could be used for various security related applications, such as network resource management, network traffic monitoring, Internet blackout and botnet-orchestrated stealth scan. We demonstrate the power of IP map to facilitate deeper insights into cyberspace with the basic case, i.e., network resource management.


The main concerns of network administrators is to have a direct and macroscopical visualization of network resources, so that they could manage them efficiently. In other words, with aid of IP map, network administrators could conduct efficient network management, including network architecture planning with limited IP resource, quick network failure locating and other security related issues. Based on the different sizes of network they manage, network administrators have the demands to visualize network resource at different granularity. For example, network carriers mainly focus on the AS-level network management and consider the resource with IP block level, while campus network administrators take care of the local area network and manage the resource at a specific IP address. Fortunately, with aid of Hilbert curve, the IP map provides the ability to show the different granularity of cyberspace by setting the order n of Hilbert curve mapping algorithm. 

\begin{figure}[t]
	\includegraphics[width=0.49\textwidth]{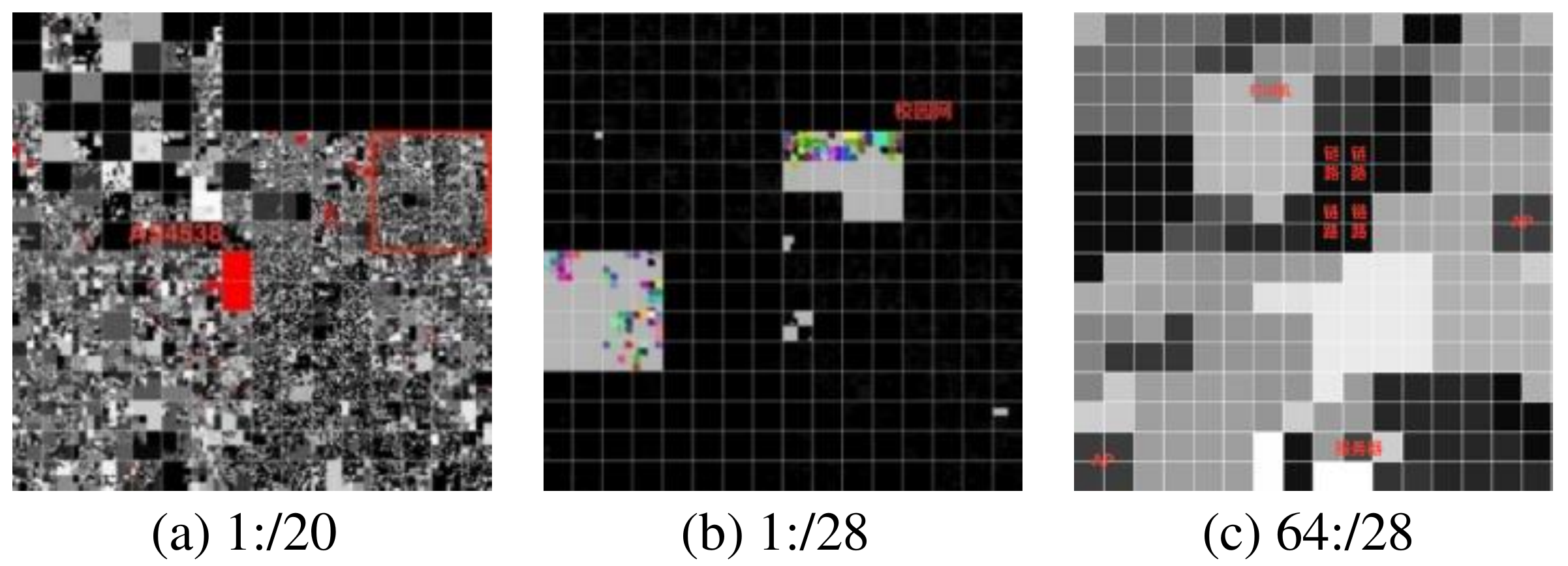}
	\centering
	\caption{The visualization of network resource from multiple scales. (a). The visualization of whole IPv4 address. (b). The visualization of one part of AS 4538. (c). The visualization of a specific campus network.}
	\label{fig_cybergis_IP}
\end{figure}

Fig. \ref{fig_cybergis_IP} depicts the visualization of network resource from multiple scales by collecting data from IANA, AS 4538 and one campus network respectively. Fig. \ref{fig_cybergis_IP}(a) depicts the visualization of whole IPv4 address. The scale is 1:/20, which means the IP prefix of /20 is aggregated into one pixel point. The network administrator in IANA could get knowledge of the IP occupancy of each organization. For example, the China Education and Research Network Center (AS 4538) contain about 2 Class A (/8) which is shown in red color. In order to further express the detailed information of AS 4538, we zoom in the CyberGIS to 1:/28 to see the details of AS 4538. As shown in Fig. \ref{fig_cybergis_IP}(b), it is obvious that the AS 4538 is divided into multiple local area networks, including backbone networks, computing center and more than 100 campus networks which is denoted with different colors. The AS level network administrator could allocated the resource according to the their demand. As for campus network administrator, they are considering more about managing the network resource at a specific IP address. Therefore, visualizing the usage of each IP address is crucial. Fig. \ref{fig_cybergis_IP}(c) presents the usage of each IP address so that network administrators could design the architecture of LAN and quickly locate the fault which may be caused by links, routers, servers, etc.

\subsection{IP-Port Map}
The IP-Port map is a 3-D visualization technique which presents the application layer of cyberspace. In addition to the basic IP management, network administrators may further concern about the detailed behavior of each host with an IP address for security supervision. So it is necessary to design an IP-Port map to provide a deep insight of each IP address. This map can be used for application layer management, such as abnormal application monitoring and application layer traffic monitoring. We demonstrate one case, i.e., application layer traffic monitoring, to show the effectiveness of CyberGIS.

\begin{figure}[t]
	\includegraphics[width=0.25\textwidth]{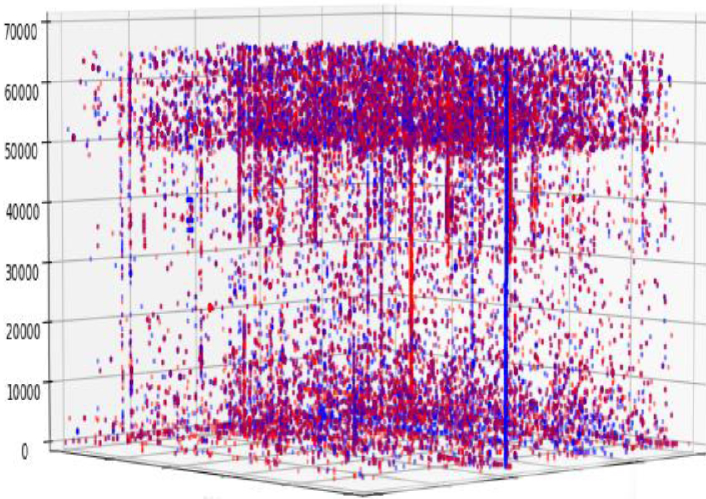}
	\centering
	\caption{The application layer traffic of a /24 IP block for a time interval. The vertical axis represent the port. The red color is the download traffic while the blue color is the upload traffic. }
	\label{fig_cybergis_IP_port}
\end{figure}

Fig. \ref{fig_cybergis_IP_port} illustrates the application layer traffic of a /24 IP block for a time interval in IP-Port map. The data is the Netflow traffic collected from campus by deploying NFDUMP in the edge routers. From the figure, a colorful visualization of network traffic is depicted and we can easily filter out the specific details of interest. An obvious visualization is that the traffic of different ports on each IP address is quite different, which is generated by the network application service based on the TCP/IP protocol. Specifically, the ports larger than 49000 or less than 1000 contains the majority of traffic. These ports belong to system ports or registered ports. Furthermore, registered ports generate more traffic than system ports which shows that IP addresses in campus are more likely to produce download behaviors. Based on the IP-Port map for traffic visualization, some network security issues such as anomalous activities and network attacks can be effectively presented. 

\subsection{AS Map}
The AS map is a two-dimensional visualization of cyberspace that provides AS-level information. Cyberspace is divided into multiple autonomous systems, which are operated according to unified routing rules and are independent between domains. And inter-AS uses inter-domain routing protocols (such as BGP) to implement traffic exchange. This type of map can be used not only for network management and security, but also for visualizing AS topology based on the characteristics of AS aggregation. The connection between the two AS nodes represents their business relationship and traffic exchange, and the topology data can come from traceroute, BGP and IRR data. As a global routing policy unit, AS defines a high-level global Internet topology for its traffic relationships, which helps us to understand the technical, economic, policy, and security requirements of a largely unregulated network peer-to-peer ecosystem. In the following, we will take the AS level topology visualization as an example to illustrate its effectiveness. 

\begin{figure}[t]
	\includegraphics[width=0.4\textwidth]{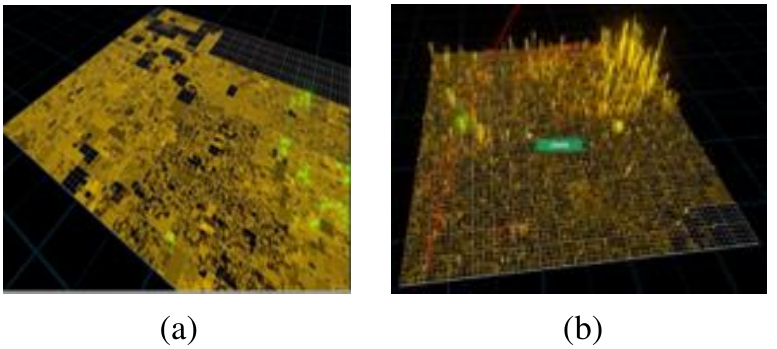}
	\centering
	\caption{The comparison between IP map and AS map. (a). The visualization of an AS for IP map. (b). The visualization of an AS and AS level topology in AS map. }
	\label{fig_cybergis_AS}
\end{figure}

Fig. \ref{fig_cybergis_AS} expresses the comparison between IP map and AS map, the data is collected from ICANN and CAIDA. As can be seen, the distribution of AS123 (green highlight) in Fig. \ref{fig_cybergis_AS}(a) is discrete due to the discontinuity of its assigned IP address blocks. It is not effective to visualize the network traffic and attack characteristics of AS in Cyberspace. While from Fig. \ref{fig_cybergis_AS}(b), we can see the AS map makes the originally dispersed IP address blocks under AS123 accumulate to the Z axis. The height of the Z axis represents the number of IP addresses which show the attribute of each AS. In addition, a colorful visualization of network traffic AS level-level topology connection is depicted and can reflect the network connectivity, assist network administrators to check hardware configuration, determine where new routes to be added, and discover bottlenecks and faults in the network, etc. Instead of representing the topological relationship using abstract points and lines, it provides the ability to describe and express in a detail and native manner compared to the map of Internet topology. At the same time, the AS backplane is fixed so that some changes in links will not affect the entire map, which also reflects the superiority of AS map. 

\section{Analogous Fourier Transform}
CyberMap presents a new perspective to visualize cyberspace. Unlike the traditional GeoMap which maps the cyberspace on geographic coordinate system (latitude and longitude), CyberMap presents a concise but effective way to provide deep insights of cyberspace. Both maps present visualization of cyberspace although they focus on different views. However, they are not conflict and will help each other instead. Similar to the famous Fourier transform which converts signals between time view and frequency view for adapting to different scenarios, each map provides unique insights into aspects of cyberspace and a combination of CyberMap and GeoMap will help to provide a more comprehensive visualization of cyberspace. Therefore, Network administrators can easily filter out the specific details of interests. 

\begin{figure}[t]
	\includegraphics[width=0.49\textwidth]{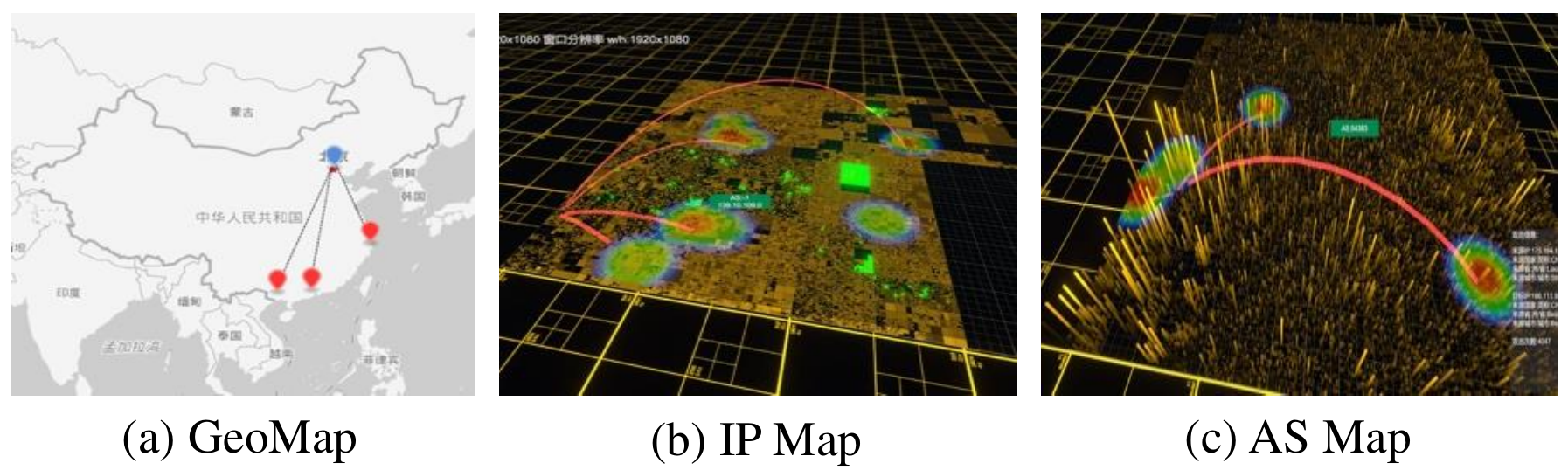}
	\centering
	\caption{A sample frame of the DDoS event on different maps. (a) visualization in GeoMap. (b) visualization in IP Map. (c) visualization in AS Map.}
	\label{comparison}
\end{figure}

Fig. \ref{comparison} presents a sample frame of the DDoS event on different maps. The data is the real world DDoS data collected from campus for one day. For each map, we divide the duration of the DDoS attack into fixed time intervals, aggregating data for each interval into a single frame. Therefore, we can get the aggregated visualization of the event. Each map provides unique aspects of DDoS event, but when they are depicted in one figure, cues from one view can illuminate the macroscopic representation of the event. Specifically, the GeoMap (Fig. \ref{comparison} (a)) presents the geographic location of the DDoS event which allow the viewer to directly get knowledge of the location of DDoS event. They are aggregated into three geographic location (the red sign). As for IP map to visualize DDoS event (Fig. \ref{comparison} (b)), we can intuitively see that the distribution of DDoS attackers (i.e. IP address) are very fragmented. Assisted by the IP map, network administrators could establish the black IP list to effectively filter out the hostile IP address. The AS map is used for locating the AS which sponsors the DDoS attack so that network administrators could reduce the reputation of that AS. A combination of each map provides multiple views of cyberspace and enables visual correlation from the macroscopic perspective. They could be coexist and the viewer could filter out the specific details of interests.

\section{Conclusion}
This paper takes first step to study the basic coordinate system of cyberspace. Firstly, we search for the space coordinate as the basic space dimension based on a number of alternative choices (such as IP address, AS numbers, etc.), and then construct 3 types of CyberCS based on the space filling curves or combination of space coordinates. Finally, we validate the effectiveness of our CyberCS and CyberMap with various real world network traces. In addition, a combination of different maps provides multiple views of cyberspace and enables visual correlation from the macroscopic perspective. Our work is the first work to study Cyberspace and there are several problems remained to be solved: (1) How to define the direction in current map? and (2) How to define the distance in current map. In the future work, we intend to do research on these problems.

\bibliographystyle{IEEEtran}

\bibliography{reference}

\end{document}